\documentclass{PoS}

\newcommand{\be}{\begin{equation}}
\newcommand{\ee}{\end{equation}}
\newcommand{\benn}{\nonumber\begin{equation}}
\newcommand{\eenn}{\nonumber\end{equation}}
\def\bea{\begin{eqnarray}} \def\eea{\end{eqnarray}}
\def\beann{\begin{eqnarray*}} \def\eeann{\end{eqnarray*}}
\def\lsim{\raise0.3ex\hbox{$<$\kern-0.75em\raise-1.1ex\hbox{$\sim$}}}
\def\gsim{\raise0.3ex\hbox{$>$\kern-0.75em\raise-1.1ex\hbox{$\sim$}}}

\title{\vspace*{-6.5cm}
{\hfill \texttt{\footnotesize CERN-PH-TH/2013-287}}
\vfill
Scale hierarchy in high-temperature QCD}

\ShortTitle{High-temperature QCD}

\author{Oscar Akerlund$^a$ and \speaker{Philippe~de~Forcrand}$^{ab}$\\
        \llap{$^a$} Institut f\"ur Theoretische Physik, ETH Z\"urich, CH-8093 Z\"urich, Switzerland\\
        \llap{$^b$} CERN, Physics Department, TH Unit, CH-1211 Geneva 23, Switzerland\\
        E-mail: \email{oscara@itp.phys.ethz.ch}, \email{forcrand@itp.phys.ethz.ch}}

\abstract{
Because of asymptotic freedom, QCD becomes weakly interacting at high 
temperature: this is the reason for the transition to a deconfined 
phase in Yang-Mills theory at temperature $T_c$. At high temperature 
$T \gg T_c$, the smallness of the running coupling $g$ induces a hierachy betwen 
the "hard", "soft" and "ultrasoft" energy scales $T$, $g T$ and $g^2 T$. This 
hierarchy allows for a very successful effective treatment where the 
"hard" and the "soft" modes are successively integrated out. However, 
it is not clear how high a temperature is necessary to achieve such 
a scale hierarchy.

By numerical simulations, we show that the required temperatures are 
extremely high. Thus, the quantitative success of the effective theory 
down to temperatures of a few $T_c$ appears surprising a posteriori.
}

\FullConference{31st International Symposium on Lattice Field Theory - LATTICE 2013\\
		July 29 - August 3, 2013\\
		Mainz, Germany}

\begin{document}

\section{Introduction}\label{sec:introduction}\noindent

Because of asymptotic freedom, QCD at high temperature $T$ is weakly coupled. At asymptotically high $T$, the theory becomes free.
The confining properties of QCD at low temperature give way to deconfinement. Also, perturbation theory becomes a good approximation.
The practical issue is how good this approximation is as a function of $T$: since the coupling runs logarithmically only, one may expect that
really high temperatures are needed. 
Indeed, the perturbative calculation of the QCD pressure $p(T)$ has represented a heroic effort, with initially rather poor
convergence properties. However, inclusion of the last perturbative term, and of the leading non-perturbative effect with a fitted
coefficient~\cite{Kajantie} results in a pressure close to available Monte Carlo data down to $T \sim {\cal O}(10 T_c)$.  A possible extension of the
perturbative approach, to include the center degrees of freedom responsible for the deconfinement transition~\cite{Kurkela}, should
extend this agreement down to a few $T_c$. 
A similarly spectacular success has been obtained with the perturbative calculation of the spatial string tension~\cite{spatial_sigma}: 
agreement with Monte Carlo data extends almost right down to $T_c$!

These extraordinary successes prompt us to examine the Debye mass, and compare perturbative predictions with Monte Carlo
data at very high temperature, in a regime where fairly good agreement is expected, but where quantitatively unknown non-perturbative
corrections should be present and can be determined.  From our study, one can estimate the temperature at which the
effective, perturbative description should become inaccurate, at least regarding the prediction of the Debye mass.

We review the perturbative approach in Sec.II, the definition of the Debye mass in Sec.III, and present our Monte Carlo results
in Sec.IV.

\section{Perturbative dimensional reduction}\label{sec:dimred}\noindent

Let us briefly recall the steps of the perturbative approach.
To implement a  finite temperature $T$, the  Euclidean time direction is compactified, with extent $1/T$. 
This renders the theory effectively 3-dimensional, for spatial distances much larger than $1/T$ or momenta $|\vec{k}| \ll T$.
Moreover, the boundary conditions in Euclidean time are periodic for bosonic, anti-periodic for fermionic fields, such that
a generic field $\phi$ admits a Fourier decomposition
\be
\tilde\phi_n = \int_0^{1/T} {\rm d}t~ \exp(i 2\pi (n+{\bf q}) t)~ \phi(x,t), ~~ {\bf q}=\{0,1/2\}
\ee
with ${\bf q}=0$ for bosons, $1/2$ for fermions. 
This implies a dispersion relation
$E_n^2 = |\vec{k}|^2 + [2\pi T (n+ {\bf q})^2 + m^2] = |\vec{k}|^2 + (m_{\rm eff}^{3d})^2$,
so that fermions acquire an effective mass ${\cal O}(\pi T)$ and decouple from the infrared physics at spatial
momenta $|\vec{k}| \ll T$. Bosons acquire an effective mass ${\cal O}(2\pi T)$ and also decouple,
except for the static mode $n=0$.

Applying this argument to QCD, one is left at high temperature with only the static modes of the gauge fields:
$\bar{A}_i \equiv A_{i,n=0}$ and $\bar{A}_0 \equiv A_{0,n=0}= \int_0^{1/T} dt~ A_0^a(\vec{x},t) \tau_a$.
The latter is related to the Polyakov loop $L$ as $L(\vec{x})=\exp(i \bar{A}_0(\vec{x}))$.

An effective, 3-dimensional action can be written in terms of the above degrees of freedom, based on
the symmetries of the theory and on a gradient expansion:
\be
S_{\rm eff}^{3d} = \int {\rm d}^3x~ [{\rm Tr} \bar{F}_{ij}^2 + m^2 \bar{A}_0^2 + (D_i \bar{A}_0)^2 + \lambda \bar{A}_0^4 + \cdots ]
\label{eq:EQCD}
\ee
which is a $3d$ Yang-Mills theory coupled to an adjoint Higgs field.
The couplings of this effective action are determined in terms of the original $4d$ Yang-Mills coupling $g(T)$ 
by matching infrared spatial correlators obtained by a perturbative expansion in the two theories. The tree-level result is $g_{\rm eff}^{3d} = g(T) \sqrt{T}$.
Note however that the $3d$ theory is non-perturbative in the infrared: $3d$ Wilson loops obey an area law,
and a $3d$ glueball spectrum exists. In fact, non-perturbative effects become manifest in a perturbative expansion,
at order $g^6(T)$ for the pressure, and the mass scale for non-perturbative excitations is $g^2(T) T$~\cite{Linde}.

Therefore, one encounters 3 different mass (or energy) scales: \\
$\bullet$ The ``hard'' scale $2\pi T$ \\
$\bullet$ The ``soft'' scale $g(T) T$ \\
$\bullet$ The ``ultra-soft'' scale $g^2(T) T$ \\
These scales are associated with the non-static modes, the electric ($\bar{A}_0$) static modes, and the magnetic ($\bar{A}_i$) static modes, respectively.
At high temperatures, $g(T)$ becomes small and these 3 scales are hierarchically separated.
Then, it makes sense to integrate them out in succession: \\
$\bullet$ Integrating out the ``hard'' scale gives the effective $3d$ theory of eq.(\ref{eq:EQCD}), which still contains the static $A_0$ mode
and is therefore called  ``Electric QCD'' (EQCD). \\
$\bullet$ Integrating out the ``soft'' scale of the $\bar{A}_0$ gives a simpler effective theory, ``Magnetic QCD'' (MQCD), whose action
is just the $3d$ Yang-Mills action, which describes the IR dynamics of EQCD, and thus of QCD. \\
In principle, the effective EQCD and MQCD actions have infinitely many local and non-local terms.
The truncation to, say, the form eq.(\ref{eq:EQCD}) is a good approximation provided that there is a scale separation between
the modes whose dynamics is preserved and those which are integrated out. \linebreak Such a hierarchy between hard, soft and ultrasoft
scales requires a small value for the coupling \nolinebreak $g(T)$.

\begin{figure}[t]
\centerline{
\includegraphics[width=0.48\linewidth]{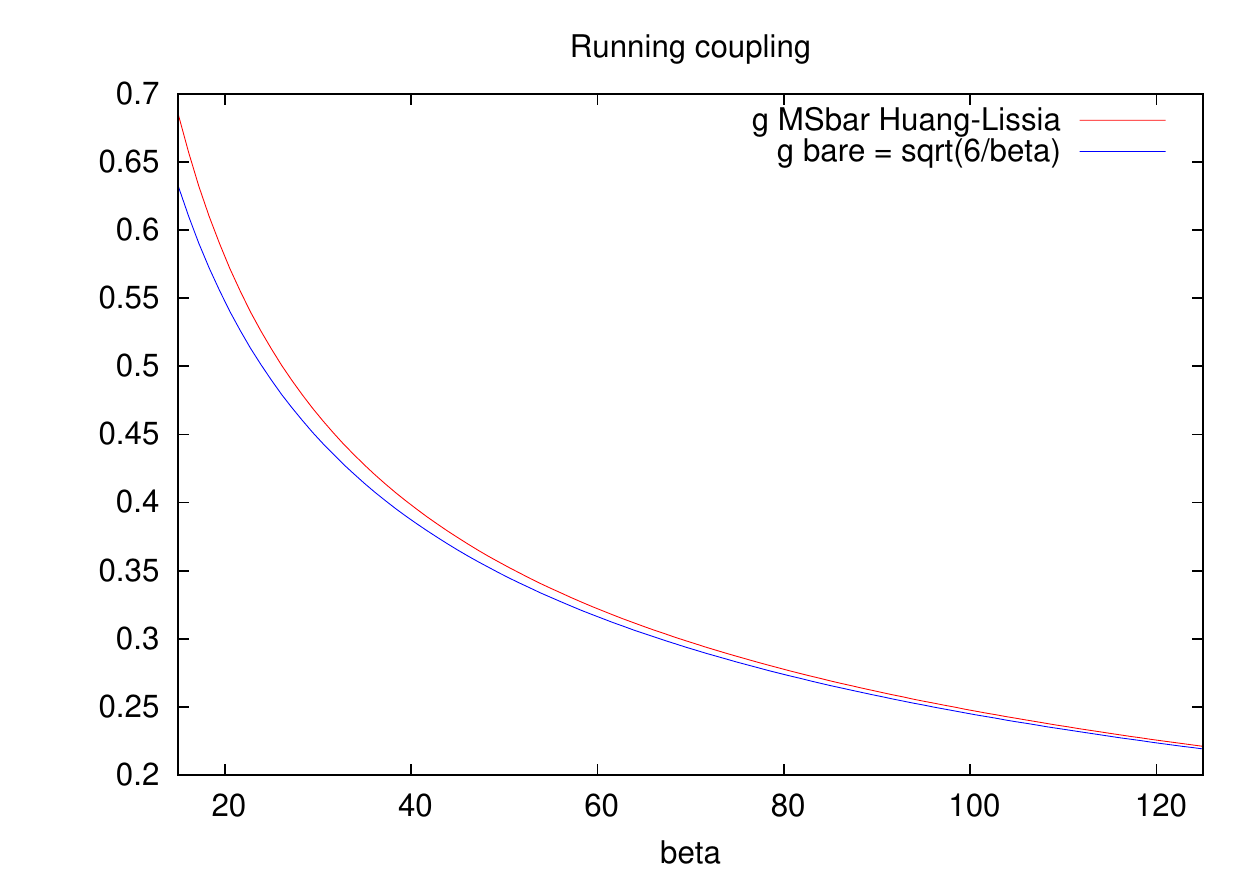}
\hspace*{0.15cm}
\includegraphics[width=0.48\linewidth,trim=0 -55 0 -55]{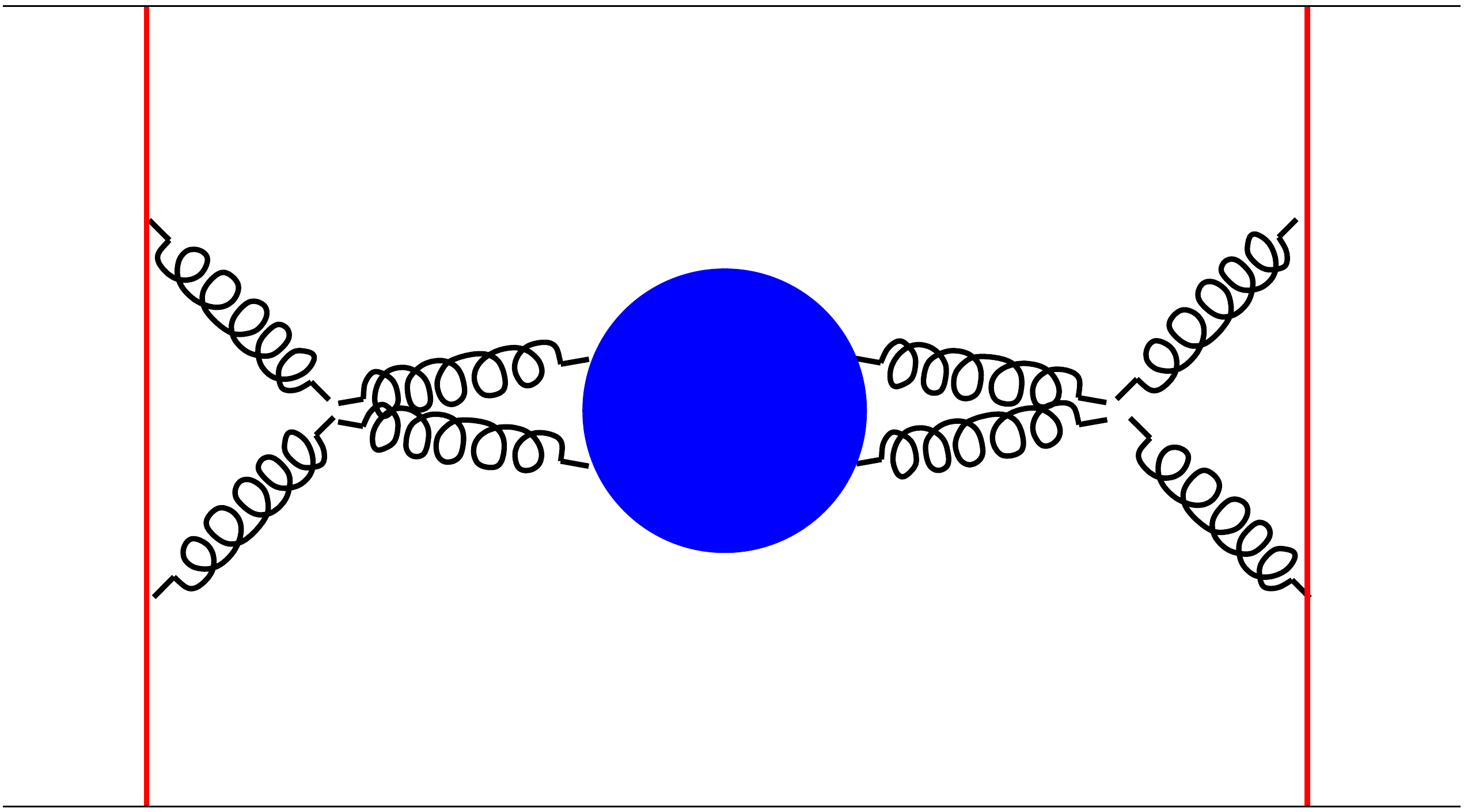}
}
\caption{{\em Left:} Running coupling as a function of the lattice coupling $\beta$, for two schemes:
the scheme of Huang and Lissia~\cite{Huang_Lissia}, favored for finite temperature, and the lattice scheme.
When the coupling is weak, the scheme dependence is mild. Note the very large values
of $\beta$ needed to reach $g \ll 1$.
{\em Right:} the 4-gluon vertex couples the $A_0$ gluons emitted by the static sources with spatial $A_i$ gluons,
themselves coupled to the non-perturbative spatial glueball channel.
}
\label{fig:running}
\end{figure}

Fig.~\ref{fig:running} {\em left} shows the running coupling $g(T)$ in the renormalization scheme of Huang and Lissia~\cite{Huang_Lissia} (MSbar scheme with 
$\mu = 4\pi T \exp(-\gamma_E - 1/22) $), popular in perturbative thermodynamic studies, where the temperature has been determined
from 2-loop running on an $N_t=3$ lattice. 
The running $g(T)$ is almost identical to the bare coupling $g$ of
the lattice action $\beta=2N_c/g^2, N_c=3$, in the regime where both are weak. Note the large values of the plaquette coupling $\beta$ required.

\section{Debye mass}\label{sec:Debye}\noindent

Consider the free energy $F_{q\bar{q}}(r)$ of two static charges $q, \bar{q}$ at distance $r$ from each other, and at finite temperature $T$.
The two charges are introduced as Polyakov loop sources $L(0)$ and $L^\dagger(r)$.
Take  the case of QED first.
The two charges interact by exchanging an $A_0$ photon. The free $A_0$ propagator is modified by $e^+ e^-$ thermal pair creation,
leading to a series $\sum_k {\cal O}(e^2)^k$ diagrams which can be resummed. As a result, the $T=0$ Coulomb
potential $1/r$ is modified at finite temperature to a Yukawa form $\frac{\exp(-m_D r)}{r}$, with the Debye mass
\be
m_D = \frac{e T}{\sqrt{3}} (1 + {\cal O}(e^2))
\ee
In QCD, the story is similar but more subtle. First, the $q\bar{q}$ free energy is obtained from the correlator $\langle {\rm Tr}L(0) {\rm Tr}L^\dagger(r)\rangle$, where ${\rm Tr}L$ is a color singlet. So, two gluons at least must be exchanged to make a singlet, making the
screening mass $2 m_E$.
The Debye mass $m_E$ is, to lowest order (noted $m_D$):
\be
m_D = gT \sqrt{\frac{N_c}{3} + \frac{N_f}{6}}
\label{eq:m_D}
\ee
for $N_c$ colors and $N_f$ quark flavors. However, at higher order the 3- and 4-gluon vertices induce a coupling of $A_0$ to $A_i$,
and thereby to the non-perturbative $3d$ glueball with mass $m_G \sim {\cal O}(g^2 T)$. This is illustrated Fig.~\ref{fig:running} {\em right}.
Therefore, $F_{q\bar{q}}(r)$ is expected to couple to both the $\bar{A}_0$ electric mode, with screening mass ${\cal O}(gT)$, and to
the $\bar{A}_i$ magnetic mode, with screening mass ${\cal O}(g^2 T)$ -- plus the non-static, ``hard'' modes with mass ${\cal O}(2\pi T)$
at short distance. This made it difficult to even define the QCD screening mass for some time.

This difficulty was resolved in \cite{Arnold_Yaffe}. The different modes behave differently under time-reversal (called ``R'' symmetry in \cite{Arnold_Yaffe}),
which changes $A_0$ to $-A_0$ and ${\rm Tr}L$ to ${\rm Tr}L^\dagger$. MQCD, not containing $A_0$, is R-even. Thus, the scale $g^2 T$ 
will be visible in R-even observables only, like ${\rm Re}{\rm Tr}L$, and the Debye mass can be {\em defined} non-perturbatively from the
asymptotic behavior of the correlator of the R-odd ${\rm Im}{\rm Tr}L$.

\begin{figure}[t]
\centerline{
\includegraphics[width=0.29\linewidth]{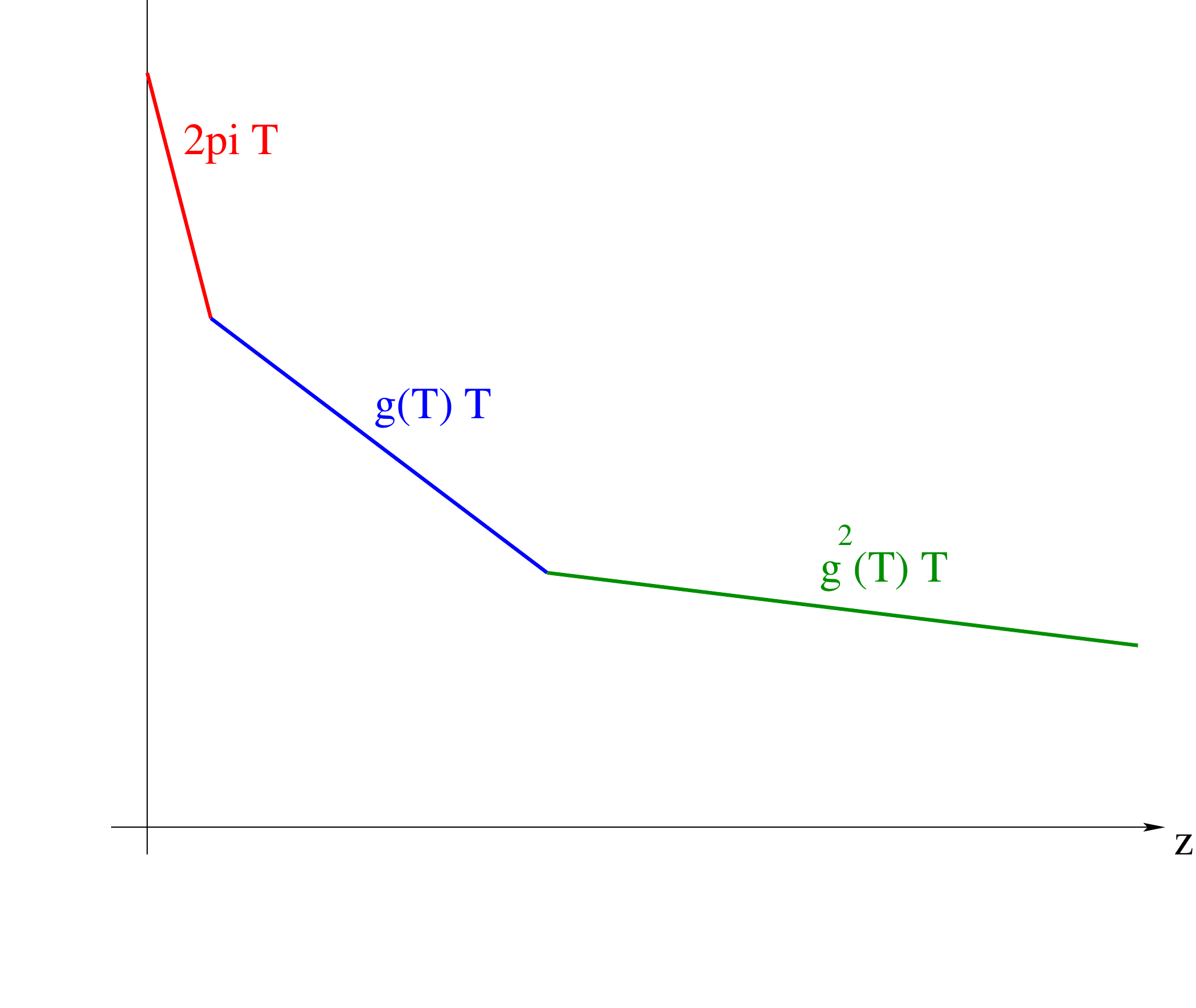}
\hspace*{0.01\linewidth}
\includegraphics[width=0.36\linewidth]{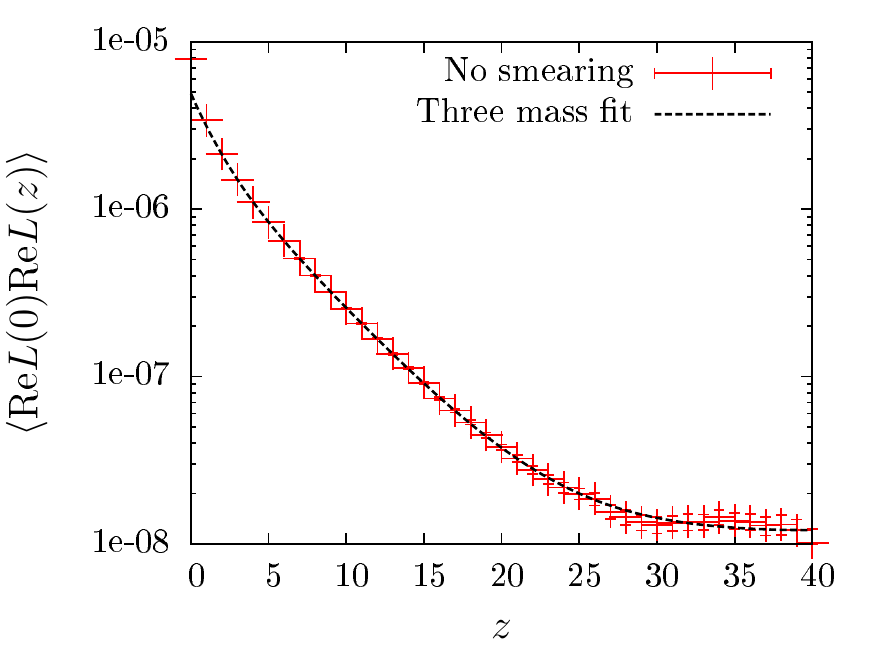}
\hspace*{-0.03\linewidth}
\includegraphics[width=0.36\linewidth]{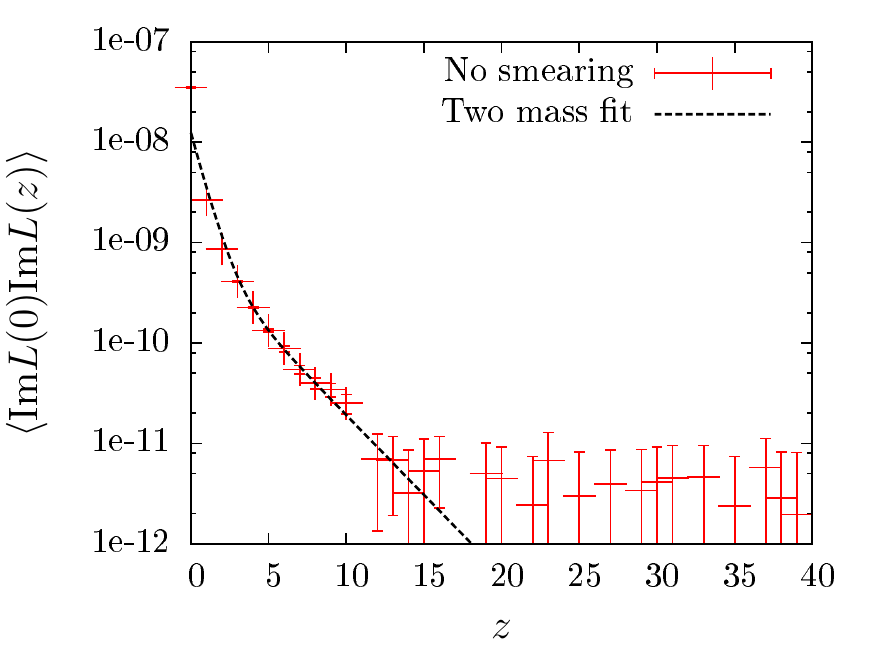}
}
\caption{{\em Left:} Schematic behaviour of the correlator $C(z)$
of the real part of the trace of the Polyakov loop (on a logarithmic scale)
as a function of separation $z$: three distinct mass scales should be
visible at high enough temperature. 
{\em Middle:} Monte Carlo data for $C(z)$ for $SU(3)$ Yang-Mills theory,
on a $24^2\times 80 \times 3$ lattice, at plaquette coupling $\beta=100$,
with a three-mass fit.
{\em Right:} Data for the correlator of the imaginary part of the Polyakov
loop trace: it is well described by a two-mass fit only. 
}
\label{fig:3scales}
\end{figure}

Thus, we separate the Polyakov loop $L$ into ${\rm Re}L \equiv \frac{L+L^\dagger}{2}$ and ${\rm Im}L \equiv \frac{L-L^\dagger}{2}$, and
measure the zero-spatial-momentum connected correlators as a function of $z$:
\bea
\langle [\sum_{xy} {\rm Tr}{\rm Re}L(x,y,0)][\sum_{xy} {\rm Tr}{\rm Re}L(x,y,z)]\rangle_c & \longrightarrow & m_{\rm eff} = \{\sim 2\pi T, ~2m_E + {\rm corr.}, ~m_G(0^{++}) \} \\
\langle [\sum_{xy} {\rm Tr}{\rm Im}L(x,y,0)][\sum_{xy} {\rm Tr}{\rm Im}L(x,y,z)]\rangle & \longrightarrow & m_{\rm eff} = \{\sim 2\pi T, ~3m_E + {\rm corr.} \}
\label{eq:C_z}
\eea
The different effective masses $m_{\rm eff}$ are parametrically separated by powers of $g(T)$. The heavier modes make a larger contribution to the 
correlator, but this contribution decays faster with $z$, leading to the behaviour sketched Fig.~\ref{fig:3scales} {\em left panel} at high temperature for the
$R$-even correlator.  These expectations are fulfilled by our Monte-Carlo data ({\em middle panel}), which is well described by a 3-mass fit where the
masses have the expected magnitude. In contrast, the $R$-odd correlator ({\em right panel}) is well described by a 2-mass fit.

Therefore, we can determine by Monte-Carlo the masses ${\cal O}(gT)$ appearing in the decay of the $R$-even and $R$-odd correlators, and 
compare them with perturbative expectations. As explained above, the masses are not simply $m_E$ because ${\rm Tr}L$ is a color singlet.
In fact, $L=\exp(i \bar{A}_0) \approx 1 + i \bar{A}_0 - \frac{1}{2} \bar{A}_0^2 - \frac{i}{6} \bar{A}_0^3 + \cdots$, with ${\rm Tr}\bar{A}_0=0$,
so that ${\rm Tr}{\rm Re}L \sim \bar{A}_0^2$ decays with mass $2 m_E$, and ${\rm Tr}{\rm Im}L \sim \bar{A}_0^3$ with mass $3 m_E$.
Finally, the leading order perturbative result for $m_E$, eq.(\ref{eq:m_D}), is modified by higher order perturbative corrections, as well
as non-perturbative ones. The latter have a known form:
 $\frac{3 g^2}{4\pi}(c \log(g) + d)$
where only $c=-1$ for the $R$-even channel is known~\cite{Mikko}. 
Given sufficiently accurate Monte-Carlo data, we can determine these 3 unknown coefficients.

\section{Monte Carlo results}\label{sec:MC}\noindent

We have measured Polyakov loop correlators eq.(\ref{eq:C_z}), for $SU(3)$ Yang-Mills theory with Wilson action, on lattices of size
$24^2\times 80 \times 3$, with plaquette coupling $\beta \in [20..120]$, corresponding to temperatures $T\in [10^{10}..10^{63}] T_c$, to make the running coupling small enough (see Fig.~\ref{fig:running} {\em left}) 
and ensure a hierarchy of scales.
Additional runs have been performed with $N_t=2$ lattices, to check for discretization errors, and with spatial size $N_x=N_y=48$,
to check for finite-size effects.

\begin{figure}[t]
\centerline{
\includegraphics[width=0.48\linewidth]{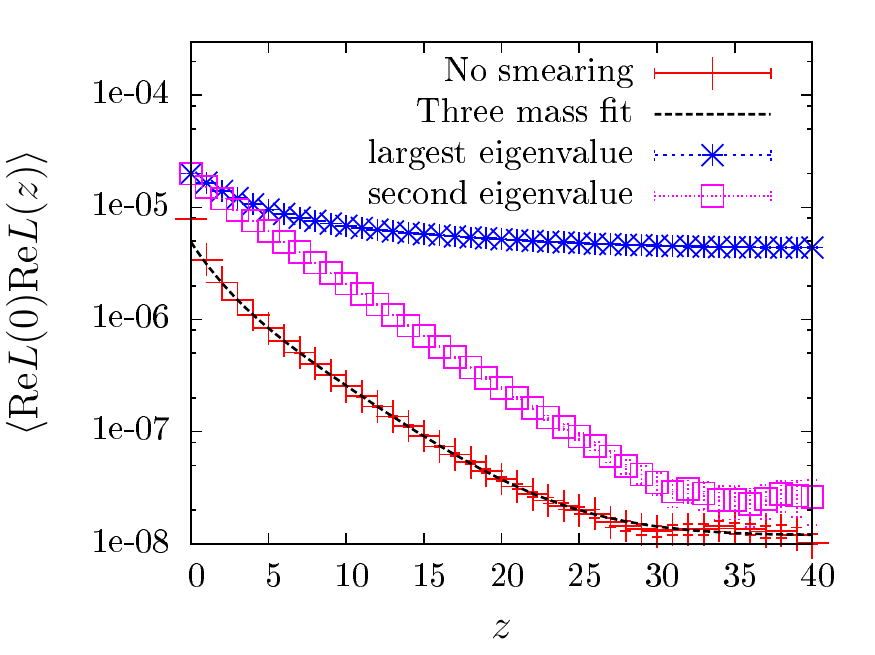}
\includegraphics[width=0.48\linewidth]{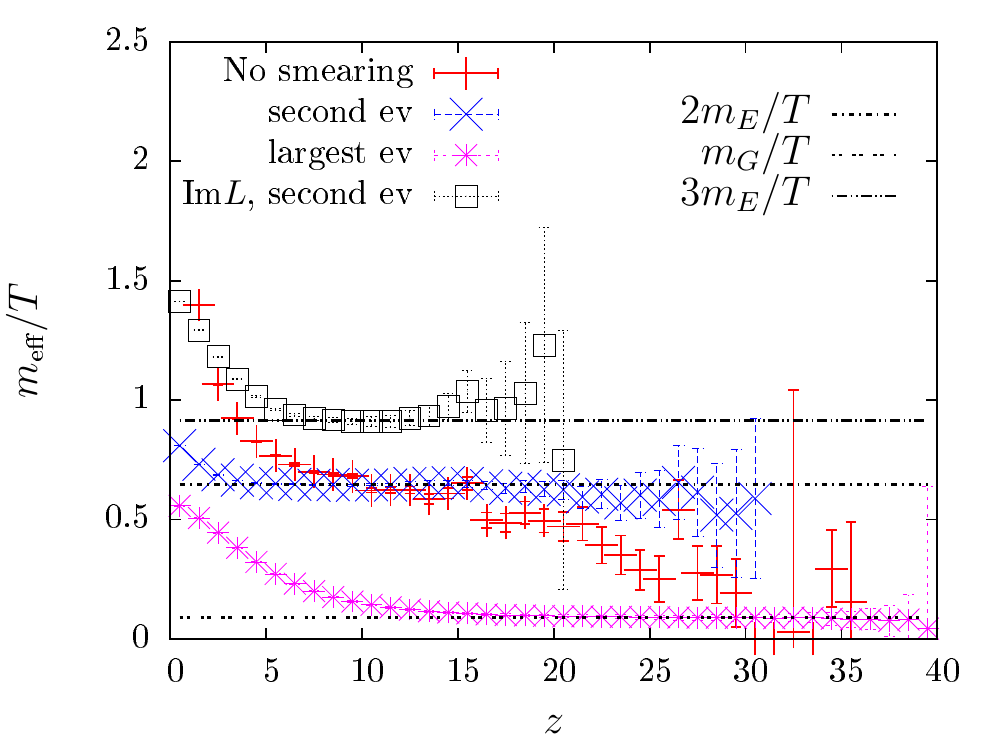}
}
\caption{{\em Left:} A clear separation of the masses is obtained by smearing
and solving a generalized eigenvalue problem~\cite{Luscher}.
{\em Right:} Effective masses as a function of $z$, for the ${\rm Re}{\rm Tr}L$
channel without (in red) and with smearing (in blue and purple): two light masses $\approx 2 m_E$ and $m_G$ are visible;
data for the smeared ${\rm Im}{\rm Tr}L$ correlator is shown in black (light mass $\approx 3 m_E$).
}
\label{fig:3masses}
\end{figure}

To suppress the hard modes, we have applied up to 160 APE smearing steps (preserving locality in $z$), obtaining for each $z$ a matrix
of correlators, indexed by the smearing level. To improve the accuracy on the screening masses, we have analyzed these matrices of 
correlators with the generalized eigenvalue method of \cite{Luscher}, where the effective mass approaches its asymptotic value
exponentially in $z$.  Fig.~\ref{fig:3masses} shows some illustrative results, for the generalized eigenvalues ({\em left}) and the resulting
effective masses ({\em right}): in the $R$-even channel, two clear mass plateaux (${\cal O}(gT)$ and ${\cal O}(g^2T)$) are obtained 
from the generalized eigenvalue approach. Without smearing these plateaux could not be determined with any confidence.

\begin{figure}[t]
\centerline{
\includegraphics[width=0.90\linewidth]{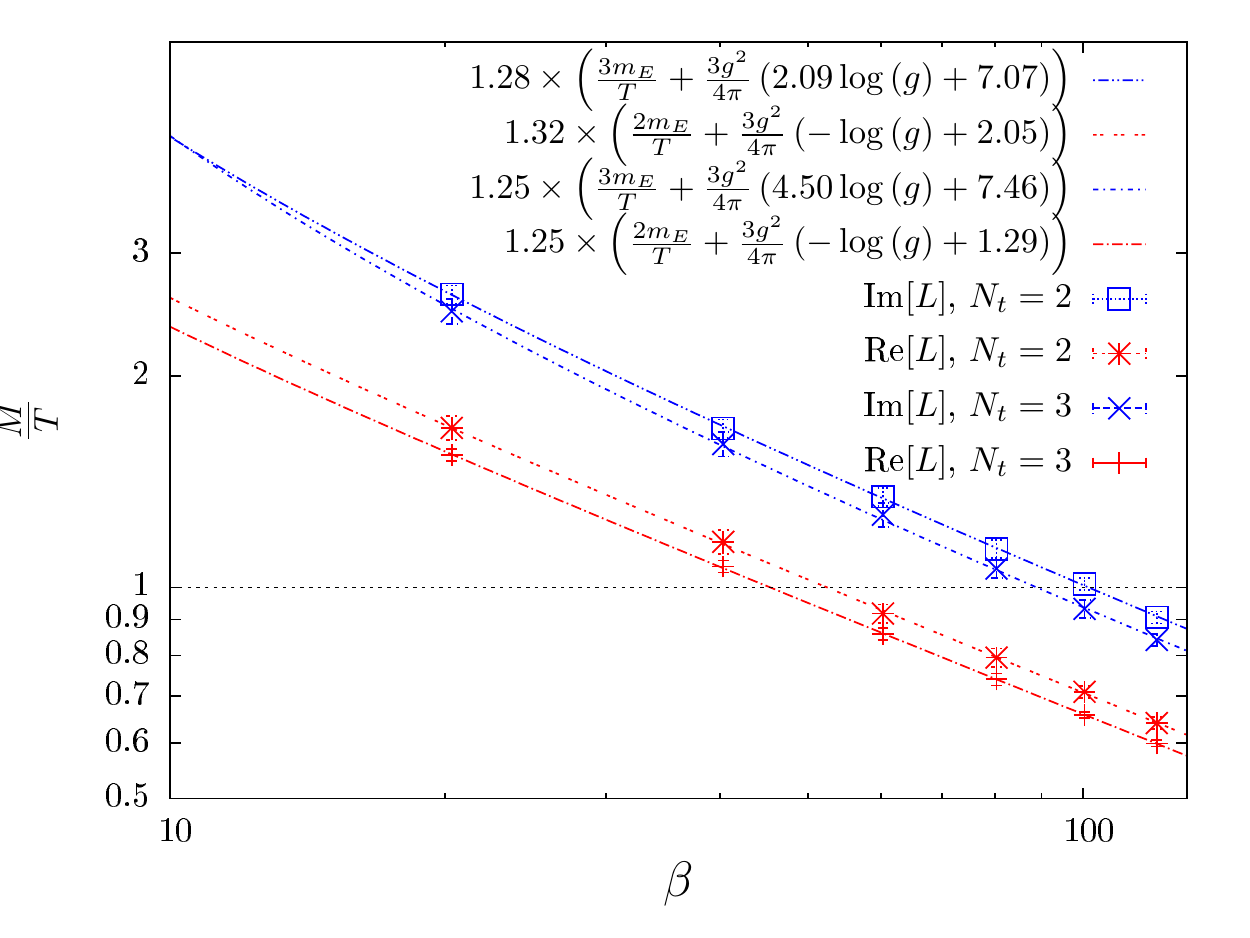}
}
\caption{Summary of our Monte Carlo measurements of Debye masses, in units
of the temperature, as a function of the plaquette coupling $\beta$,
on $N_t=2$ and $N_t=3$ lattices. 
The masses $\approx 2 m_E$ (in red) and $\approx 3 m_E$ (in blue) 
characterizing the decays 
of correlators of $A_0^2$ and $A_0^3$, respectively, are shown. The fits
include leading non-perturbative corrections to the perturbative prediction.
}
\label{fig:Debye_summary}
\end{figure}

Our Debye mass measurements are presented in Fig.~\ref{fig:Debye_summary}.  The dimensionless ratio $M/T$ is shown as a function
of the plaquette coupling $\beta$, for both $R$-even and $R$-odd channels and for $N_t=3$ and $2$ lattices.
In these coordinates, the parametric dependence $M \sim gT$  becomes $M/T \sim \beta^{-1/2}$, which is why we use logarithmic 
scales. Indeed, the data fall on nearly straight lines. A small systematic deviation, similar for both channels, is visible between 
$N_t=3$ and $N_t=2$ data. This is a discretization error, which is smaller for $N_t=3$ and $\sim 25\%$ there. 
Otherwise, the data are nicely consistent with the perturbative plus non-perturbative contributions discussed Sec.~\ref{sec:Debye}.
Our precision allows to determine the unknown non-perturbative coefficients with reasonable accuracy.

This success should not prevent us from mentioning difficulties: some are technical, some fundamental.
On the technical side, we have noticed that smearing of the links entering the Polyakov loops does not conserve $R$-parity:
our two observables ${\rm Re}L$ and ${\rm Im}L$ are no longer purely $R$-even and $R$-odd, respectively. In practice however,
allowing for mixing in our analysis produced entirely consistent results. More seriously, 
we have had problems with the glueball mass, ${\cal O}(g^2 T)$, which governs the $z$-asymptotics of the
$R$-even correlator. It has been determined in the $3d$ theory in \cite{Teper}. In our $(3+1)d$ setup, we find it a factor 2-3
 lighter than expected. Its accurate determination is challenging, because finite-size effects are significant,
and because a noisy disconnected part must be subtracted from the $R$-even correlator. Moreover, the glueball may mix with the
$x$- or $y$-flux states carried by the $x$- and $y$-oriented Polyakov loops, which have similar masses.

The fundamental difficulty is rather a puzzle. The horizontal line $M/T=1$ in Fig.~\ref{fig:Debye_summary} is crossed in the
mid-range of the -- very weak -- couplings which we have considered. On the left side of the figure, the Debye masses are
close to $\pi T$, and cannot be cleanly separated from the hard modes. Similarly, the glueball mass, which varies as $1/\beta$,
faster than the $1/\sqrt{\beta}$ of the Debye mass, rapidly becomes heavier towards the left edge of the figure.
In fact, when $T \sim 2 T_c$, the mass hierarchy $2\pi T \gg m_E \gg m_G$ is inverted to $m_E\sim 2\pi T < m_G$, as reported, e.g., in \cite{Hart}.
This should make the predictions of perturbative EQCD grossly inaccurate at such temperatures, contrary to empirical observations~\cite{spatial_sigma}. 
Our study validates the perturbative dimensional reduction approach at temperatures circa $10^{10} T_c$ and above.
Its success at much smaller temperatures can be considered a ``lucky surprise'', lessening the appeal of a more elaborate
treatment as in \cite{Kurkela}.


\begin{thebibliography}{99}
\bibitem{Kajantie}
  K.~Kajantie, M.~Laine, K.~Rummukainen and Y.~Schroder,
  Phys.\ Rev.\ D {\bf 67}, 105008 (2003)
  [hep-ph/0211321].

\bibitem{Kurkela} 
  A.~Kurkela,
  Phys.\ Rev.\ D {\bf 76}, 094507 (2007)
  [arXiv:0704.1416 [hep-lat]];
  P.~de Forcrand, A.~Kurkela and A.~Vuorinen,
  Phys.\ Rev.\ D {\bf 77}, 125014 (2008)
  [arXiv:0801.1566 [hep-ph]].

\bibitem{spatial_sigma}
  Y.~Schroder and M.~Laine,
  PoS LAT {\bf 2005}, 180 (2006)
  [hep-lat/0509104].

\bibitem{generic}
  K.~Kajantie, M.~Laine, K.~Rummukainen and M.~E.~Shaposhnikov,
  Nucl.\ Phys.\ B {\bf 458}, 90 (1996)
  [hep-ph/9508379].

\bibitem{Linde}
  A.~D.~Linde,
  Phys.\ Lett.\ B {\bf 96}, 289 (1980).

\bibitem{Huang_Lissia}
  S.~-z.~Huang and M.~Lissia,
  Nucl.\ Phys.\ B {\bf 438}, 54 (1995)
  [hep-ph/9411293].

\bibitem{Arnold_Yaffe}
  P.~B.~Arnold and L.~G.~Yaffe,
  Phys.\ Rev.\ D {\bf 52}, 7208 (1995)
  [hep-ph/9508280].

\bibitem{Mikko}
  M.~Laine and M.~Vepsalainen,
  JHEP {\bf 0909}, 023 (2009)
  [arXiv:0906.4450 [hep-ph]].

\bibitem{Luscher}
  M.~Guagnelli {\it et al.}  [ALPHA Collaboration],
  Nucl.\ Phys.\ B {\bf 535}, 389 (1998)
  [hep-lat/9806005].

\bibitem{Teper}
  M.~J.~Teper,
  Phys.\ Rev.\ D {\bf 59}, 014512 (1999)
  [hep-lat/9804008];
  B.~Lucini and M.~Teper,
  Phys.\ Rev.\ D {\bf 66}, 097502 (2002)
  [hep-lat/0206027].

\bibitem{Hart}
  A.~Hart, M.~Laine and O.~Philipsen,
  Nucl.\ Phys.\ B {\bf 586}, 443 (2000)
  [hep-ph/0004060].

\end{thebibliography}
\end{document}